\documentclass{ifacconf}

\usepackage{graphicx}      %
\usepackage{natbib}        %
\usepackage{physics}
\usepackage{amssymb,amsmath,xcolor,graphicx,xspace,colortbl,rotating} %
\usepackage{textcomp}
\usepackage{amsmath,mathtools}  %

\newtheorem{lemma}{Lemma}
\newtheorem{property}{Property}
\DeclareMathOperator{\diag}{diag}

\begin{document}
\setlength{\parskip}{2pt}
\begin{frontmatter}

\title{Recursive identification with regularization and on-line hyperparameters estimation}

\author[First]{Bernard Vau} 
\author[Second]{Tudor-Bogdan Airimitoaie} 

\address[First]{Exail, 12 avenue des Coquelicots, 94385 Bonneuil sur Marne, France (e-mail: bernard.vau@exail.com).}
\address[Second]{Univ. Bordeaux, CNRS, Bordeaux INP, IMS, 33405 Talence, France (e-mail: tudor-bogdan.airimitoaie@u-bordeaux.fr)}

\begin{abstract}                %
This paper presents a regularized recursive identification algorithm with simultaneous on-line estimation of both the model parameters and the algorithms hyperparameters. A new kernel is proposed to facilitate the algorithm development. The performance of this novel scheme is compared with that of  the recursive least squares algorithm in simulation.
\end{abstract}


\end{frontmatter}
\section{Introduction}
Recursive identification is unavoidable whenever one needs a model estimation performed on-line (real-time). The principle dates back to (\cite{Whitaker58}) and is almost as old as the concept of identification itself. Recursive identification algorithms have been designed for the three classes of classical identification schemes: Prediction Error Methods (PEM) (\cite{Ljung83}), Instrumental Variable (\cite{Young11}), and Pseudo-Linear Regression (PLR) (\cite{LandauSpringer11}) (for which some improvements  have been made recently (\cite{Vau21})). For several decades since the famous paper of (\cite{Astrom65}), identification  has been dominated, directly or indirectly, by the Maximum Likelihood (ML) approach, and this is also the case for recursive schemes. On the contrary, Bayesian concepts consisting in introducing a prior knowledge in estimation, did not received a significant attention, in spite of earlier attempts to estimate dynamical systems  (\cite{Leamer72}), (\cite{Akaike79}).

It is only since the beginning of the 2010s, that the Bayesian approach has attracted a considerable interest in the identification field with the emergence of kernel-based identification schemes,  coming from Machine Learning (\cite{Pillonetto10}). By applying some regularization techniques, namely modifying a least squares criterion, the bias-variance trade-off can be significantly improved, especially if the initial problem is ill conditioned (\cite{Pillonetto14}). The most significant kernels embedding a prior knowledge of intrinsic systems properties (such as stability and smoothness of the impulse response) are the Stable-Spline (SS) Kernel (\cite{Pillonetto10}), the Diagonal Correlated (DC) and the Tuned-Correlated (TC) kernels (\cite{Chen12}). These kernels are tuned with a number of hyperparameters whose values can be determined beforehand from data, using a Hierarchical Bayes approach (see \cite[p. 107]{Pillonetto22}). So far, to the authors knowledge, this kernel approach has been employed in a recursive context in (\cite{Prando16}), (\cite{Romeres16}) and (\cite{Illg22}).

In the sequel one presents a novel recursive and regularized identification scheme for the estimation of the impulse response of a discrete Linear Time Invariant (LTI) system, and where the model is a Finite Impulse Response (FIR) filter. The proposed scheme is more easily implementable online than the one disclosed in (\cite{Romeres16}). Like in this last reference, it is assumed that the prior is not known at initialization, which requires to estimate online the prior hyperparameters at the same time as the parameters. In general, in regularized schemes, the number of Kernel hyperparameters  is reduced (see \cite{Pillonetto22}). Here, in order to make the algorithm implementable on line, one proposes unusually a prior with a number of hyperparameters equal to the number of parameters, the prior structure (smoothness of the impulse response and stability) being imposed by putting some constraints to those hyperparameters. This prior is finally very close to the TC prior (\cite{Chen12}). The online estimation of the hyperparameters aims at minimizing the cost of the marginal likelihood criterion (Empirical Bayes approach) \cite[chap.4]{Pillonetto22}, by using a recursive gradient with projection, while the parameters are estimated with an algorithm quite similar to recursive least squares (RLS).

The paper is structured as follows: Section 2 shows how the recursive least-squares can be implemented in case of regularized criterion, following a Bayesian approach, Section 3 proposes a prior structure inspired from the TC prior, Section 4 details the algorithm mixing on-line parameters and hyperparameters estimation. At last, Section 5 displays some simulations showing the performances of this novel algorithm.

\section{Problem statement}
\subsection{Classical propagation equations for recursive estimation}
Let us consider the ``true" discrete-time system $G_0(q)$, which is assumed to be stable and proper, ($q$ being the forward shift operator). The sequences $\{u(t)\}$,  $\{y(t)\}$, $\{e(t)\}$ correspond respectively to the input, output and a centred gaussian white noise. This noise has a variance $\sigma^2$ assumed to be known. One has 
\begin{equation}
y(t)=G_0(q)u(t)+e(t)
\label{eq1}
\end{equation}

\noindent The model is a finite impulse filter $G$ with
\begin{equation}
G(q)=\sum_{k=1}^{n}b_kq^{-k}
\label{eq2}
\end{equation}

\noindent Our purpose is to estimate the vector $\theta$, where \[ \theta^T=[b_1 \; b_2 \; \cdots b_n] \] 

\noindent The regularized least-squares criterion with data up to time $t$  denoted as $J_0(t)$ is
\begin{equation}
J_0(t)=\sum_{i=1}^{t}\left( y(i)-\phi^T(i-1)\hat{\theta}(t)\right)^2+\hat{\theta}^{T}(t)\sigma^2\Pi_0^{-1}\hat{\theta}(t),
\nonumber %
\end{equation}
\noindent $\hat{\theta}(t)$ being the estimation of $\theta$ at time $t$, and \[\phi(t)=[u(t-1),~u(t-2),\ldots u(t-n)]^T \]  

The positive definite matrix $\Pi_0$ is the regularization matrix (corresponding to the kernel functions in a Machine Learning approach). By cancelling out the criterion gradient with respect to the vector $\hat{\theta}$, one finds immediately

\begin{equation}
\hat{\theta}(t)=\left( \sum_{i=0}^{t-1}\phi(i)\phi(i)^T+\sigma^2\Pi_0^{-1} \right)^{-1}\sum_{i=1}^{t}\phi(i-1)y(i)
\label{eq2b}
\end{equation}

\noindent Let us introduce the positive definite matrix $F$ as
\begin{equation}\label{eq_F_Pi}
	F^{-1}(t)= \sum_{i=0}^{t-1}\phi(i)\phi(i)^T+\sigma^2\Pi_0^{-1}
\end{equation}

\noindent Its propagation equations are
\begin{align}
{F^{-1}(t+1)=F^{-1}(t)+\phi(t)\phi^{T}(t)} ,~
{F^{-1}(0)=\sigma^2\Pi_0^{-1}}
\label{eq3}
\end{align}

\noindent By using the matrix inversion  lemma, one has classically (\cite{LandauSpringer11}, p.64) 
\begin{equation}
F(t+1)=F(t)-\frac{F(t)\phi(t)\phi^{T}(t)F(t)}{1+\phi^{T}F(t)\phi(t)}
\label{eq3b}
\end{equation}

\noindent On the other hand, it is well known (see \cite{LandauSpringer11}, p.63) that the propagation of $\hat{\theta}$ is given by

\begin{subequations}\label{eq4}
\begin{align}
{\hat{\theta}(t+1)=\hat{\theta}(t)+F(t)\phi(t)\frac{\varepsilon^{o}(t+1)}{1+\phi^T(t)F(t)\phi(t)}} \\
{\varepsilon^{o}(t+1)=y(t+1)-\phi^{T}(t)\hat{\theta}(t)}
\end{align}
\end{subequations}

\noindent In a Bayesian perspective, the prior knowldege is incorporated in $\Pi_0$ (which corresponds to the covariance matrix of this  prior). Therefore, in a recursive estimation algorithm, incorporating this prior is equivalent to specify $F(0)$. However, when one performs an identification in real-time, in general one cannot use some available data so as to specify this prior (by applying for example the Hierarchical Bayes approach) allowing to determine the optimal prior hyperparameters, by maximizing a marginal likelihood function. It is the reason why, we consider in this paper the issue of computing on-line the regularisation matrix $\Pi(t)$ in parallel with the computation of $\hat{\theta}(t)$.

\subsection{Case where the regularization matrix is no longer invariant}

Let us assume that by a specific estimation algorithm (which will be detailed below), the regularization matrix $\Pi$ is updated at each sample time, yielding $\Pi(t)$. We assume that $\Pi$ is expressed from a vector of hyperparameters $\eta$ which is no longer constant. From \eqref{eq_F_Pi} and \eqref{eq3}, one can write now
\begin{align}
F^{-1}(0,\eta(0))&=\sigma^2\Pi^{-1}(0)\\
F^{-1}(t,\eta(t))&=\sum_{i=0}^{t-1}\phi(i)\phi(i)^T+\sigma^2\Pi^{-1}(t)\\
F^{-1}(t+1,\eta(t+1))&=\sum_{i=0}^{t}\phi(i)\phi(i)^T+\sigma^2\Pi^{-1}(t+1)
\label{eq5c}
\end{align}

\noindent  The propagation equation of $F^{-1}$ becomes
\begin{equation}
{F^{-1}(t+1,\eta(t))=F^{-1}(t,\eta(t))+\phi(t)\phi^T(t)} 
\label{eq6a}
\end{equation}

\noindent On the other hand, from \eqref{eq4}, one has
\begin{equation}
\hat{\theta}(t+1,\eta(t))=\hat{\theta}(t,\eta(t))+\frac{F(t,\eta(t))\phi(t)\varepsilon^{o}(t+1)}{1+\phi^T(t)F(t,\eta(t))\phi(t)}
\label{eq6c}
\end{equation}
An update of $F$ and $\hat{\theta}$ in function of $\eta(t+1)$ is detailed in Section~\ref{sec_4_1}.

\subsection{Marginal Likelihood in a recursive context}
Set
\[  Y^{T}(t+1)=\begin {bmatrix} y(1) \; y(2)  \cdots y(t+1) \end{bmatrix} \]
In the hierarchical Bayes approach, the vector of hyperparameters $\eta$ is estimated from data at first, by maximizing the Marginal Likelihood function $L(\eta|Y)$.  Afterwards the \textit{maximum a posteriori (MAP) estimate} $\hat{\theta}=argmax \; \textbf{p}(\theta|Y)$ is computed, $\textbf{p}$ being the probability density function (pdf). 
In a recursive context, where data are obtained on-line, and available up to time $t+1$, the maximization of the Marginal Likelihood function $L(\eta(t+1)|Y(t+1))$ and the computation of the MAP $\hat{\theta}(t+1)=argmax \; \textbf{p}(\theta(t+1)|Y(t+1), \eta(t+1))$ must be computed alternatively, so as to take into account only available data at time $t+1$. For this purpose set
\[ \Phi(t)=\begin{bmatrix}  \phi(0) &  \phi(1) & \cdots & \phi(t) \end{bmatrix}^T  \]
\[  \Sigma(t+1)=\Phi(t) \Pi(t+1) \Phi^{T}(t)+\sigma^2I_{t+1}  \]

\noindent assuming that the prior expectation of $\theta$ is null, and that $\theta$, $\eta$ are normally distributed, the marginal LogLikelihood function $log \; L(\eta(t+1),Y(t+1)$   is  (\cite{Pillonetto22}, p.108)
\begin{multline}
\log L(\eta(t+1) |Y(t+1))=-\frac{1}{2}\log(2\pi \left|\Sigma(t+1) \right|)\\
-\frac{1}{2}Y^{T}(t+1)\Sigma^{-1}(t+1)Y(t+1)
\label{eq12}
\end{multline}

\noindent 0ne has the following result:
\begin{lemma}
\label{lem1}
The derivative of $\log L(\eta(t+1) |Y(t+1))$ with respect to the $k$ entry of $\eta(t+1)$ (denoted as $\eta_k(t+1)$)   is
\begin{multline}
\frac{\partial log(L(t+1))}{\partial \eta_k(t+1)}=\frac{1}{2}Tr \left( \left[\left( \Pi(t+1)-\right.\right.\right.\\
\left.\left.\left.-\hat{\theta}(t+1)\hat{\theta}^{T}(t+1) \right)-\sigma^2F(t+1)\right]
\frac{\partial  \Pi^{-1}(t+1)}{\partial \eta_k(t+1)} \right)
\label{eq17}
\end{multline}
\end{lemma}
Proof: See Appendix A.\hfill

\section{Prior structure and hyperparameters estimation}

\subsection{Prior structure}

One proposes a matrix regularization structure inspired from the TC Kernel (\cite{Chen12}) but where $\eta$ has $n$ entries so as to facilitate the on-line estimation. This is contrary to the usual method where the amount of hyperparameters is generally very reduced (see for example \cite{Pillonetto22}). Afterwards some constraints on the hyperparameters are introduced so as to incorporate the prior knowledge associated to the system impulse response (stability, exponential decay). The regularization matrix is defined as
\begin{equation}
 \Pi(t,\eta)=UW(t,\eta)U^T 
\label{eq18}
\end{equation}

\noindent where $U$ is upper triangular with every nonzero element equal to one and $W(t,\eta)=\diag(\exp(\eta(t)))$,
 with $\eta^{T}(t)=[\eta_1(t) \; \eta_2(t) \; \cdots \eta_n(t)]$.

\noindent Now, some constraints are imposed to $\eta$.
\begin{itemize}
\item Constraint C1: $\eta_2(t)<\eta_1(t) \; \; \forall \; t$ \\
\item Constraint C2: $\eta_k(t)-2\eta_{k+1}(t)+\eta_{k+2}(t)=0$
\end{itemize}

\noindent Note that the vector subspace of $\mathbb{R}^n$ subject to constraints C1 and C2 is convex, moreover these constraints are equivalent to imposing $\eta_k(t)=\eta_1(t)-(k-1)\alpha(t)$, where $\alpha(t)>0$. One has also 
\[ W_{i,i}(t,\eta)=\exp(-\alpha(t)) W_{i-1,i-1}(t,\eta) \]
\noindent with  $0<\alpha(t)$. This prior structure differs from the TC kernel only  because of the term $W_{n,n}$ (see \cite[eqs.~(20)--(22)]{Carli17}). As the TC kernel, the proposed one is well suited to damped systems. One can now perform a change of basis where the parameters estimate is $\hat{\theta}^{'}$ and the regressor $\phi^{'}(t)$ such that 
\begin{equation}
{\hat{\theta}^{'}(t)=U^{-1}\hat{\theta}(t)},\quad
{\phi^{'}(t)=U^T\phi(t)},
\label{eq21a}
\end{equation}
\begin{multline}
{F^{'-1}(t+1,\eta(t+1))=U^{-1}F^{-1}(t+1,\eta(t+1))U^{-T}=}\\
{\sum_{i=0}^{t}\phi^{'}(i)\phi^{'T}(i)+\sigma^2 W^{-1}(t+1,\eta(t+1))}
\label{eq21c}
\end{multline}

\noindent and in this basis from \eqref{eq2b}, one can write
\begin{multline}
{\hat{\theta}^{'}(t+1,\eta(t+1))=} \\
{\left( \sum_{i=0}^{t}\phi^{'}(i)\phi^{'}(i)^T+\sigma^2 W^{-1}(\eta(t+1)) \right)^{-1}\sum_{i=0}^{t}\phi^{'}(i)y(i)}
\label{eq21d}
\end{multline}

\noindent Thus, in the basis where $\hat{\theta}^{'}$ is the estimated parameters vector and $\phi^{'}$ the regressor, the associated prior $W$ is nothing but a kernel matrix $K^{'}$ corresponding to the DI Kernel (see \cite{Chen12}), with
\begin{equation}
 K_{i,j}^{'}=\beta \lambda^{i}\delta(i-j) 
\label{eq21e}
\end{equation}

\noindent where $\delta(i)=1$  if $i=0$  and  $\delta(i)=0$  otherwise, and $\beta>0$. \\

\begin{property}
\label{prop1}
The prior $\Pi(t,\eta)$ as defined in \eqref{eq18} under the constraints C1 and C2 is stable.
\end{property}
\noindent Proof: One verifies immediately that $\Pi_{k,k}^{1/2}(t,\eta) \leq \frac{W_{1,1}}{\sqrt{1-\lambda}}\lambda^{\frac{k-1}{2}}$, hence  $\lim_{n \to \infty} \sum_{k=1}^{n}\Pi^{1/2}_{k,k}  (t,\eta) \leq \frac{W_{1,1}}{\sqrt{1-\lambda }}\frac{1}{1-\lambda^{1/2}}$, \\ thus $\lim_{n \to \infty} \sum_{k=1}^{n}\Pi^{1/2}_{k,k}(t,\eta) < \infty$, which is the condition for the prior to be stable (see Lemma 5.1 of \cite{Pillonetto22}).\hfill$\Box$

\begin{property}
	The prior $\Pi(t,\eta)$ as defined in \eqref{eq18} has the property of maximum entropy.
\end{property}
The proof follows \cite[Section~IV]{Carli14} and is omitted. %

\subsection{On-line hyperparameters estimation}

The estimation procedure of the vector $\eta$ consists in finding $\eta$ that minimizes $J_h(t)=-\log (L(\eta(t) |Y(t)))$.  By combining \eqref{eq17}, \eqref{eq21a}--\eqref{eq21c}, one can write (the dependence with respect to $t$ being omitted for sake of simplicity) 

\begin{align}
\frac{\partial J_h}{\partial \eta_k}&=\frac{1}{2} Tr\left( U \left(W-\hat{\theta}^{'}\hat{\theta}^{'T}-\sigma^{2}F^{'}\right)U^{T}U^{-T}\frac{\partial W^{-1}}{\partial \eta_k}U^{-1}\right)
\label{eq22}\\
\frac{\partial J_h}{\partial \eta_k}&=\frac{1}{2} Tr\left( \left(W-\hat{\theta}^{'}\hat{\theta}^{'T}-\sigma^{2}F^{'}\right)\frac{\partial W^{-1}}{\partial \eta_k}\right)
\label{eq23}
\end{align}

\noindent and finally
\begin{equation}
\frac{\partial J_h}{\partial \eta_k}=\frac{1}{2} \left( 1-\frac{\hat{\theta_k}^{'2}+\sigma^2F^{'}_{kk}}{e^{\eta_k}}\right).
\label{eq24}
\end{equation}

\noindent Set $\Psi(t)$ such that
\begin{equation}
\Psi(t)=\begin{bmatrix} 
             1-\frac{\hat{\theta}_1^{'2}(t)+\sigma^2F^{'}_{11}(t)}{e^{\eta_1(t)}}\\
						 1-\frac{\hat{\theta}_2^{'2}(t)+\sigma^2F^{'}_{22}(t)}{e^{\eta_2(t)}}\\
						   \vdots \\
						 1-\frac{\hat{\theta}_n^{'2}(t)+\sigma^2F^{'}_{nn}(t)}{e^{\eta_n(t)}}
						\end{bmatrix}
\label{eq25}
\end{equation}

\noindent One defines $\Psi_p(t)$ as the projection of $\Psi(t)$ onto the vector subspace subject to constraints C1 and C2 with
\[\Psi_p(t)=Proj (\Psi(t)), \]

\noindent where the function $\Psi \mapsto Proj(\Psi(t))$ is such that
\begin{subequations}
\begin{align}
{\Psi_p^{*}=\left( I_n-C^{T}\left(CC^T\right)^{-1}C\right)\Psi(t)} \\
{\text{if} \; \; \Psi_{p(1)}^{*}(t)<\Psi_{p(2)}^{*}(t) \; \; \; \Psi_p(t)=\Psi_p^{*}(t)}\\
{\text{else} \; \; \Psi_{p(k)}=\Psi^{*}_{p(1)}- (k-1)\epsilon},
\end{align}
\label{eq26}
\end{subequations}
\noindent $\epsilon>0$ is close to $0$, and 
\[C=\begin{bmatrix}
            1 & -2 & 1 & 0 & 0 & \cdots & 0 & 0 &0 \\
						0 &  1 & -2 & 1 & 0 & \cdots & 0 & 0 &0 \\
						\vdots & \vdots &  \vdots &  \vdots &  \vdots &  \vdots &  \vdots &  \vdots & \vdots \\
						0 & 0 & 0 & 0 & 0 & \cdots & 1 & -2 & 1 \\
						\end{bmatrix} \]
						
\noindent the size of C being $(n-2\times n)$. The estimation on-line of $\eta$ is therefore performed using a gradient descent with projection ($\gamma>0$ being the corresponding adaptation gain)
\begin{equation}
\eta(t+1)=\eta(t)-\gamma\Psi_p(t).
\label{eq27}
\end{equation}

\section{Overall algorithm}

\subsection{Update of the adaptation gain an estimated parameters vector}\label{sec_4_1}

By combining \eqref{eq5c} and \eqref{eq6a}, one obtains
\begin{multline}
{F^{-1}(t+1,\eta(t+1))=F^{-1}(t+1,\eta(t))+}\\
{\sigma^2\left(\Pi^{-1}(\eta(t+1))-\Pi^{-1}(\eta(t))\right)} .
\label{eq6b}
\end{multline}
\and Introducing \eqref{eq27}, one gets
\begin{multline}
F^{'-1}(t+1,\eta(t+1))-F^{'-1}(t+1,\eta(t))=\\
=\sigma^{2}\left( W^{-1}(\eta(t+1))-W^{-1}(\eta(t))\right).
\label{eq28}
\end{multline}
\noindent For a square matrix $X$ one has $\partial X=-X\partial X^{-1} X$, and one can write
\begin{multline}
F^{'}(t+1,\eta(t+1))=F^{'}(t+1,\eta(t))-\sigma^{2} F^{'}(t+1,\eta(t))\cdot\\
\cdot\left( W^{-1}(\eta(t+1))-W^{-1}(\eta(t))\right) F^{'}(t+1,\eta(t))
\label{eq28b}
\end{multline}
\noindent Note that \eqref{eq28b} results from a first order approximation which can be made more accurate by using a truncated Neumann series expansion of higher order. On the other hand, from \eqref{eq21d}
\begin{align*}
\hat{\theta}^{'}(t+1,\eta(t+1))&=F^{'}(t+1,\eta(t+1))\sum_{i=0}^{t}\phi^{'}(t)y(i+1)\\
\hat{\theta}^{'}(t+1,\eta(t))&=F^{'}(t+1,\eta(t))\sum_{i=0}^{t}\phi^{'}(t)y(i+1)
\end{align*}
\noindent and by combining with \eqref{eq28b}, one gets
\begin{multline}
\hat{\theta}^{'}(t+1,\eta(t+1))= \left(I_n-\sigma^{2} F^{'}(t+1,\eta(t))\cdot\right.\\
\left. \cdot\left( W^{-1}(\eta(t+1))-W^{-1}(\eta(t))\right) \right)\hat{\theta}(t+1,\eta(t))
\end{multline}

\subsection{Summary}

The algorithm computing alternatively the estimation of parameters and hyperparameters is given by\footnote{Remark: since there is no explicit matrix inversion in the update of $F$, the algorithm's computational complexity is $O(n^2)$ at each sampling instant. Note also that the matrix inversion in~\eqref{eq12} has complexity $O(t^3)$.}
\begin{subequations}
\begin{multline}
\hat{\theta}'(t+1,\eta(t))= \hat{\theta}'(t,\eta(t))+\\
+F'(t,\eta(t))\phi'(t)\frac{\varepsilon^{o}(t+1)}{1+\phi^{'T}(t)F'(t,\eta(t))\phi^{'}(t)}
\end{multline}
\begin{multline}
F^{'}(t+1,\eta(t))=F^{'}(t,\eta(t))-\\
-\frac{F^{'}(t,\eta(t))\phi'(t)\phi^{'T}(t)F'(t,\eta(t))}{1+\phi^{'T}F'(t,\eta(t))\phi'(t)}
\end{multline}
\begin{equation}
{\eta(t+1)=\eta(t)-\gamma\Psi_p(t)}
\end{equation}
\begin{multline}
F^{'}(t+1,\eta(t+1))=F^{'}(t+1,\eta(t))-\sigma^{2} F^{'}(t+1,\eta(t))\cdot\\
\cdot\left( W^{-1}(\eta(t+1))-W^{-1}(\eta(t))\right) F^{'}(t+1,\eta(t))
\end{multline}
\begin{multline}
	\hat{\theta}^{'}(t+1,\eta(t+1))= \left(I_n-\sigma^{2} F^{'}(t+1,\eta(t))\cdot\right.\\
	\left. \cdot\left( W^{-1}(\eta(t+1)-W^{-1}(\eta(t))\right) \right)\hat{\theta}(t+1,\eta(t))
\end{multline}
\end{subequations}

\section{Simulation results}
\begin{figure}
	\centering
	\includegraphics[width=\columnwidth]{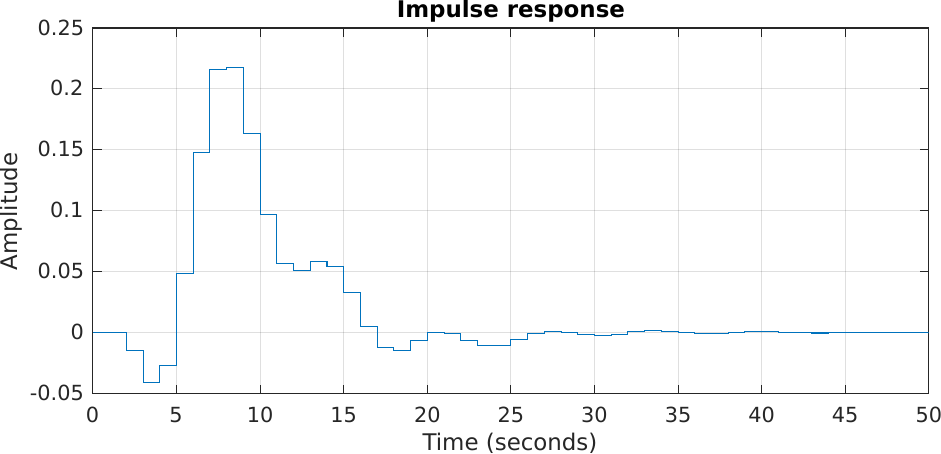}
	\caption{Impulse response of $G_o$.}
	\label{fig_impulse_Gz}
\end{figure}

\begin{figure}[h!]
	\centering
	\includegraphics[width=\columnwidth]{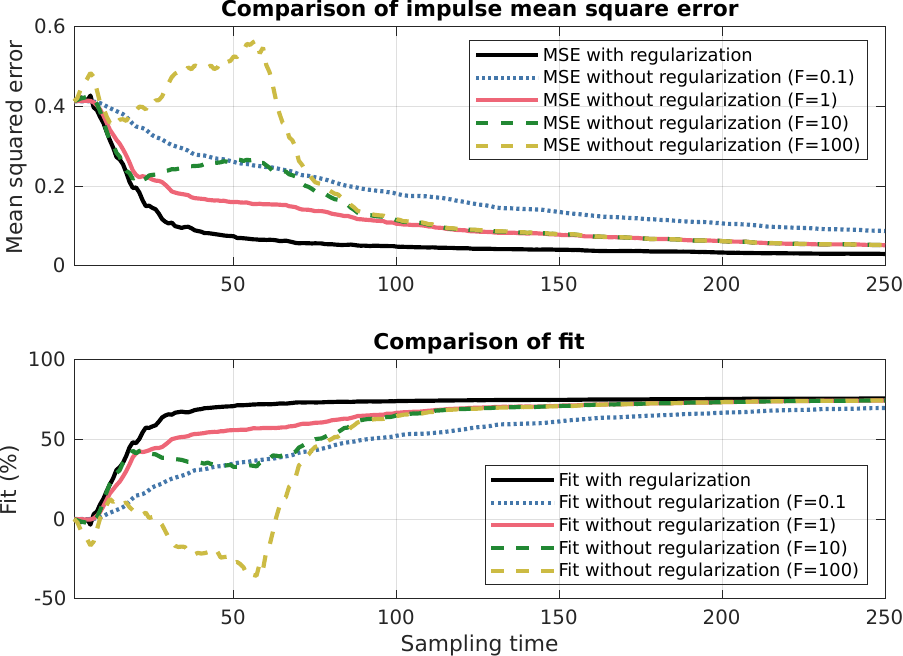}
	\caption{Comparison with recursive least squares with and without regularization ($\eta_1(0)=\log(0.001)$).}
	\label{fig_comp_fit_mse}
\end{figure}

In this section, the proposed regularization based recursive identification algorithm is compared with the classic recusrive least squares (RLS) method (see \cite[Chapter~5]{LandauSpringer11}).  The identification data is generated using the nominal model:
\begin{multline}\label{eq_no_model}
	G_o(z)=\frac{0.02008 z^{-1} + 0.04017 z^{-2} + 0.02008 z^{-3}}{1 - 1.561 z^{-1} + 0.6414 z^{-2}}\cdot\\
	\cdot \frac{-0.7334 z^{-1} + 1.516 z^{-2} - 0.7591 z^{-3} + 0.6941z^{-4}}{1 - 1.282 z^{-1} + 1.298 z^{-2} - 0.4757 z^{-3} + 0.1775z^{-4}}\nonumber
\end{multline}
The impulse response of the transfer function $G_o$ is shown in Fig.~\ref{fig_impulse_Gz}.
The input is 250 samples from a zero mean random gaussian signal with 0.5 standard deviation. Measurement noise is added to the output of the model in the form of zero mean gaussian signal with standard deviation 0.05. The number of parameters is $n=50$. The initial gain matrix $F$ of the RLS algorithm is chosen diagonal, with identical diagonal values. Four RLS identifications have been done, with diagonal element given by 0.1, 1, 10, or 100. Figure~\ref{fig_comp_fit_mse} shows the comparison with the regularized RLS algorithm proposed in this paper. For each algorithm, the curves are obtained by averaging over 10 simulations. The average signal-to-noise ratio for these 10 simulations is 12.3 dB (computed using the standard deviation). The identification results depend on the initialization of the hyperparameters. The following initialization of $\eta(t)$ is used here:
\begin{equation}
	\eta_k(0)=\begin{cases}
		\log(0.001),&\text{ if }k=1,\\
		\log(0.9)+\eta_{k-1}(0),&\text{ for } k>1.
	\end{cases}
\end{equation}
The estimation gain $\gamma$ is equal to 1. At each time-step when a new input-output data pair is available, the identified model of each algorithm is updated and the impulse response is computed. The mean square error (MSE) between the impulse response of the identified and the nominal models are shown in the upper plot of Figure~\ref{fig_comp_fit_mse}. Similarly, after each new iteration of the recursive algorithms, the fit of the model to the identification data is computed and the result for the various algorithms are shown in the lower plot of Figure~\ref{fig_comp_fit_mse}.

\begin{figure}[h!]
	\centering
	\includegraphics[width=\columnwidth]{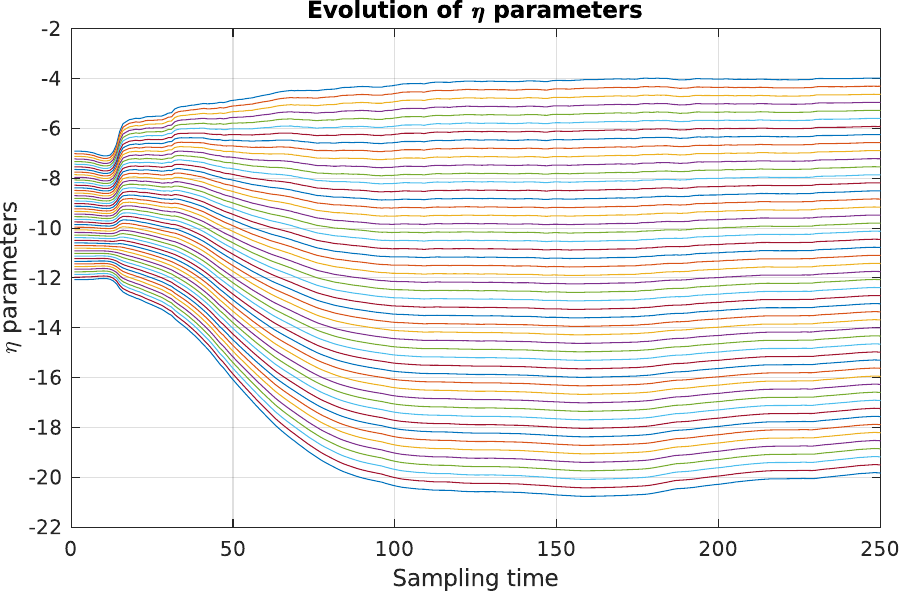}
	\caption{Evolution of the hyperparameters ($\eta_1(0)=\log(0.001)$).}
	\label{fig_evol_eta}
\end{figure}

\begin{figure}
	\centering
	\includegraphics[width=\columnwidth]{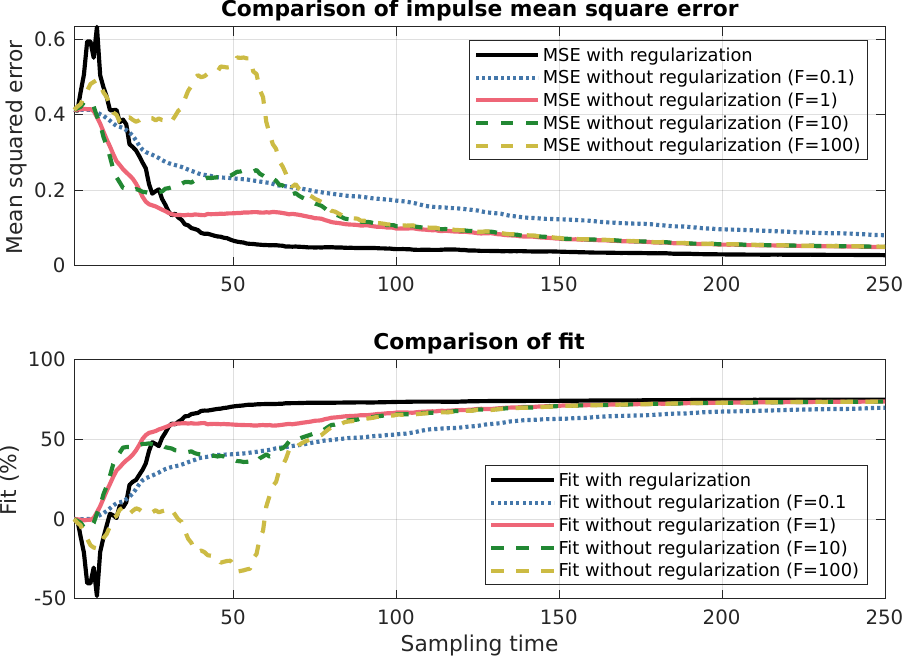}
	\caption{Comparison with recursive least squares with and without regularization ($\eta_1(0)=\log(0.1)$).}
	\label{fig_comp_fit_mse0_1}
\end{figure}

\begin{figure}
	\centering
	\includegraphics[width=\columnwidth]{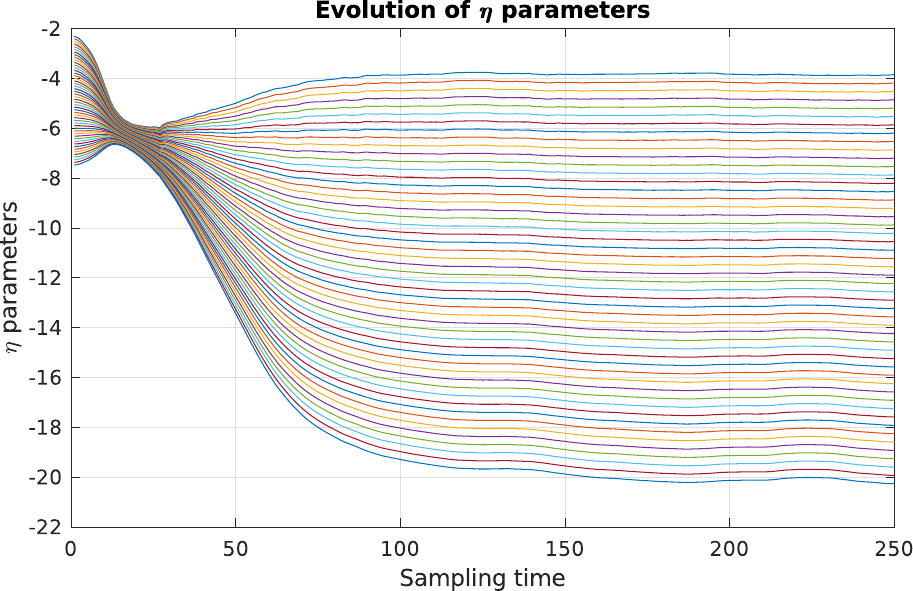}
	\caption{Evolution of the hyperparameters ($\eta_1(0)=\log(0.1)$).}
	\label{fig_evol_eta0_1}
\end{figure}

For the computation of the fit, the full input-output data is used, even though the algorithms only use this data only one sample at a iteration time. Clearly, for both MSE and Fit, the regularized least squares algorithm yields better results. Figure~\ref{fig_evol_eta} shows the evolution of the hyperparameters in one simulation. Figure~\ref{fig_comp_fit_mse0_1} shows the comparison based on an average over 10 simulations for a different initialization of the hyperparameters using $\eta_1(0)=\log(0.1)$. The evolution of the hyperparameters in one of the simulations is displayed in Figure~\ref{fig_evol_eta0_1}. Despite a degradation of the results during the first 15~s, the regularized least squares solution reaches the optimal solution more rapidly than the other algorithms. This is related also to the reach of the optimal hyperparameter values.

\section{Concluding remarks}
A novel recursive identification algorithm with online estimation of model parameters and regularization matrix hyperparameters has been presented. A simulation comparison with classical recursive least squares shows the superiority of the proposed method during the initial instants of the recursive algorithm, and the faster reduction of the prediction error squares. Future work will focus on the initialization and the convergence properties of the hyperparameters estimation.

\bibliography{biblio}             %

\appendix
\section{Proof of Lemma \ref{lem1}}  
\noindent By using the inversion lemma, one has
\begin{align}
\Sigma^{-1}(t+1)&=\frac{1}{\sigma^2}I_{t+1}-\frac{1}{\sigma^2}\Phi(t)\left(\sigma^2\Pi^{-1}(t+1)+\right.\nonumber\\
&\quad\left.+\Phi^T(t)\Phi(t)\right)^{-1}\Phi^{T}(t)
\end{align}

\noindent Since $|\Sigma^{-1} |=1/|\Sigma |$, and by using the Sylvester theorem about determinants, one gets (the dependence with respect to $t$ is omitted for sake of clarity)

\begin{equation}
\left | \Sigma^{-1} \right|= \left | \left( \sigma^2\Pi^{-1}+\Phi^{T}\Phi \right)^{-1} \right | |\sigma^2\Pi^{-1} |  \left |\frac{I_N}{\sigma^2} \right |
\end{equation}
$N$  being the data number (This number doesn't play any role in the sequel), thus
\begin{equation}
log \left | \Sigma^{-1} \right|=-Nlog (\sigma^2)-log \left | \sigma^2\Pi^{-1}+ \Phi^{T}\Phi \right |+ log \left | \sigma^2\Pi^{-1} \right |
\end{equation}

\noindent Note that $\frac{\partial log \left| X(z) \right|}{\partial z}=Tr\left( X^{-1}\frac{\partial X}{\partial z} \right)$, yielding ($\eta_k$ is the $k$th entry of $\eta$)
\begin{multline}
\frac{\partial log \left| \Sigma^{-1} \right|}{\partial \eta_k}=-Tr \left[ \left(\sigma^2 \Pi^{-1}+\Phi^{T}\Phi \right)^{-1}\frac{\partial \sigma^2 \Pi^{-1}}{\partial \eta_k} \right]\\
+Tr  \left[\sigma^{-2}\Pi \frac{\partial \sigma^2\Pi^{-1}}{\partial \eta_k} \right]
\end{multline}

\noindent but $\Phi^{T}(t)\Phi=\sum_{i=0}^{t}\phi(i)\phi^{T}(i)$, and therefore $(\sigma^2\Pi^{-1}(t+1)+\Phi^{T}(t)\Phi(t))^{-1}=F(t+1)$, therefore one gets

\begin{multline}
\frac{\partial log \left | \Sigma^{-1}(t+1) \right |}{\partial \eta_k(t+1)}=\\
=Tr \left [ \left(\sigma^{-2}\Pi(t+1)-F(t+1)\right)\frac{\sigma^2\partial \Pi^{-1}(t+1)}{\partial \eta_k(t+1)} \right ]
\label{eq_partial_sigma_m1}
\end{multline}

\noindent Now let us consider the expression $l(t+1)=Y^{T}(t+1)\Sigma^{-1}(t+1)Y(t+1)$. One obtains
\begin{multline}
   l=\frac{1}{\sigma^2}Y^{T}(t+1)Y(t+1)-\frac{1}{\sigma^2}Y^{T}(t+1)\Phi(t)\left(\sigma^{2}\Pi^{-1}(t+1)\right.\\
   \left.+\Phi^{T}(t)\Phi(t) \right)^{-1}\Phi^{T}(t)Y(t+1)
\end{multline}
Consequently we get
\begin{multline}
\frac{\partial l(t+1)}{\partial \eta_k(t+1)}=-\sigma^{-2}Y^{T}(t+1) \cdot\\
\cdot\Phi \frac{\partial \left( \sigma^{2}\Pi^{-1}(t+1)+\Phi^{T}(t)\Phi(t) \right)^{-1}}{\partial \eta_k(t+1)} \Phi^{T}(t)Y(t+1)
\end{multline}

\noindent Since $\frac{\partial X^{-1}(z)}{\partial z}=-X^{-1}\frac{\partial X}{\partial \eta_k} X^{-1}$, one has
\begin{multline}
\frac{\partial l(t+1)}{\partial \eta_k(t+1)}=\sigma^{-2}\sum_{i=1}^{t}y(i+1)\phi^{T}(i)F(t+1) \cdot\\
\cdot\frac{\partial \sigma^2\Pi^{-1}}{\partial \eta_k(t+1)} F(t+1)\sum_{i=1}^{t}\phi(i)y(i+1)
\end{multline}

\noindent now from \cite[Eq. (3.36)]{LandauSpringer11} , $F(t+1)\sum_{i=1}^{t}\phi(i)y(i+1)=\hat{\theta}(t+1)$, therefore
\begin{equation}
\frac{\partial l(t+1)}{\partial \eta_k(t+1)}=\sigma^{-2}\hat{\theta}^{T}(t+1)\frac{\partial \sigma^2\Pi^{-1}}{\partial \eta_k(t+1)} \hat{\theta}(t+1)
\label{eq_partial_l}
\end{equation}

By combining \eqref{eq_partial_sigma_m1} and \eqref{eq_partial_l}, one obtains the gradient of  $log L(\eta(t+1) |Y(t+1))$
\begin{multline}
\frac{\partial log(L(t+1))}{\partial \eta_k(t+1)}=\frac{1}{2}\left(Tr \left[ \left(\sigma^{-2}\Pi(t+1)-F(t+1)\right)\right.\right.\\
\left.\left.\frac{\partial  \sigma^2\Pi^{-1}(t+1)}{\partial \eta_k(t+1)} \right]-\frac{1}{\sigma^2}\hat{\theta}^{T}(t+1)\frac{\partial \sigma^2 \Pi^{-1}(t+1)}{\partial \eta_k(t+1)} \hat{\theta}(t+1)\right)
\end{multline}
and therefore
\begin{multline}
\frac{\partial log(L(t+1))}{\partial \eta_k(t+1)}=\frac{1}{2}Tr \left[ \left[\sigma^{-2}\left( \Pi(t+1)-\right.\right.\right.\\
\left.\left.\left.-\hat{\theta}(t+1)\hat{\theta}^{T}(t+1) \right)-F(t+1)\right]
\frac{\partial  \sigma^2\Pi^{-1}(t+1)}{\partial \eta_k(t+1)} \right]
\label{eq_likelihood2}
\end{multline}

\end{document}